\documentclass{article}

\usepackage[style=authoryear, doi=false,isbn=false,url=false, eprint=false, citestyle=numeric-comp, firstinits, sorting=none]{biblatex}
\usepackage[utf8]{inputenc}
\usepackage{amsmath}
\usepackage{hyperref}
\PassOptionsToPackage{hyphens}{url}
\usepackage[toc,page]{appendix}
\usepackage{graphicx}
\usepackage{amsfonts}
\usepackage[left=1in,top=1in,right=1in,bottom=1in]{geometry}
\usepackage{setspace}
\usepackage{xcolor}
\usepackage{tabularx}
\usepackage{makecell}

\newcolumntype{Y}{>{\raggedright\arraybackslash}X}
\title{Blurring cluster randomized trials and observational studies: Two-Stage TMLE for sub-sampling, missingness, and few independent units}

\author{Joshua R. Nugent, Carina Marquez, Edwin D. Charlebois, Rachel Abbott, Laura B. Balzer \\for the SEARCH Collaboration}

\date{August 2022}

\begin{document}

\maketitle

\begin{abstract}
\noindent
Cluster randomized trials (CRTs) often enroll large numbers of participants; yet due to resource constraints, only a subset of participants may be selected for  outcome assessment, and those sampled may not be representative of all cluster members. Missing data also present a challenge: if sampled individuals with measured outcomes are  dissimilar from those with missing outcomes, unadjusted  estimates of arm-specific endpoints and the intervention effect may be biased. Further, CRTs often enroll and randomize few clusters, limiting statistical power and  raising concerns about finite sample performance. Motivated by SEARCH-TB, a cluster randomized trial aimed at reducing incident tuberculosis infection, we demonstrate interlocking methods to handle these challenges. First, we extend Two-Stage targeted minimum loss-based estimation (TMLE) to account for three sources of missingness: (1) sub-sampling; (2) measurement of baseline status  among those sampled; and (3) measurement of final status among those in the incidence cohort (persons known to be at risk at baseline). Second, we critically evaluate the assumptions under which sub-units of the cluster can be considered the conditionally independent unit, improving precision and statistical power but also causing the CRT to behave  like an observational study. Our application to SEARCH-TB highlights the real-world impact of different assumptions on measurement and dependence;
estimates relying on unrealistic assumptions suggested the intervention increased the incidence of TB infection by 18\% (risk ratio [RR]=1.18, 95\% CI: 0.85-1.63), while estimates accounting for the sampling scheme, missingness, and within community dependence found the intervention decreased the incident TB by 27\% (RR=0.73, 95\% CI: 0.57-0.92).
\end{abstract}


\section{Introduction}
\label{intro}

In randomized controlled trials, the intervention is sometimes randomized to groups of participants rather than to individuals \cite{hayes_cluster_2017, campbell_how_2014, donner_design_2010, eldridge_practical_2012}. For example, it would be impractical to evaluate a new teaching method by randomizing students within classrooms, but much more feasible if classrooms were randomized. In group or cluster randomized trials (CRTs), correlation between the outcomes within a cluster may arise due to shared environmental factors, shared exposure to the intervention (or control), and interactions between individuals within a cluster.
This dependence violates the common regression assumption that all observations are independent and identically distributed (i.i.d.), complicating statistical estimation and inference.

A number of well-established methods can account for the dependence of observations in a cluster  \cite{liang_longitudinal_1986, fitzmaurice_applied_2012, hayes_cluster_2017}. However, not all methods can address practical challenges arising in CRTs. First, outcomes may not be measured on all participants in each cluster. This could occur by design, for example, if measurement of a rare or expensive outcome only occurred in a sub-sample of participants. Failing to adjust for sampling can result in biased point estimates and misleading inference \cite{horvitz_generalization_1952, robins_new_1986, laan_targeted_2011}. Additionally, incomplete ascertainment of outcomes among all (or the selected subset of) participants can bias results if the outcomes are not missing completely at random (MCAR) \cite{rubin_inference_1976, robins_analysis_1995}. Individuals whose outcomes are not measured are likely different than those who were fully observed; for example, students who are absent on an exam day may be systematically different than those present. If this systematic missingness is influenced by the intervention (for example, a new teaching technique  improves motivation and attendance, influencing exam scores and the probability of measurement), the risk of bias is even larger. This is a common problem: a recent review found that missing data were present in 93\% of CRTs, 55\% of which simply performed a complete-case analysis \cite{fiero_statistical_2016}.  

Second, resource constraints often limit the number of clusters in CRTs. Indeed, a review of 100 CRTs  found 37\% with fewer than 20 clusters \cite{kahan_increased_2016} and another review of 100 CRTs found a median of 33 clusters \cite{selvaraj_characteristics_2013}. Further, in CRTs with many clusters, key subgroup analyses might be conducted within strata defined by cluster-level covariates (e.g., region), limiting the number of randomized units included in that analysis. As the number of clusters shrinks, chance imbalance on covariates that influence the outcome becomes more likely. Accounting for these covariates and other outcome predictors can increase precision (e.g. \cite{fisher_statistical_1932, tsiatis_covariate_2008,  moore_covariate_2009, hayes_cluster_2017, benitez_defining_2022}). However, in analyses with few clusters, including too many covariates can lead to overfitting, and it is often not clear which covariates to select for optimal performance \cite{balzer_adaptive_2016}.

Third, statistical inference often relies on (i) tests with known finite sample properties that may be inefficient or (ii) the asymptotic behavior of estimators that may not hold in CRT analyses with a limited number of clusters. For example, generalized estimating equations (GEE) and generalized linear mixed models (GLMMs) are two common approaches for analyzing CRTs \cite{laird_random-effects_1982, liang_longitudinal_1986}; both rely on having a ``sufficient'' number of clusters. The exact recommendation varies, with some suggesting GEE can be used with as few as 10 clusters \cite{pan_small-sample_2002}, while others suggest that these approaches (without small-sample corrections) should be avoided without 30 or more clusters \cite{kreft_introducing_1998, hayes_cluster_2017, murray_design_2018}. Altogether, inference based on a small number of clusters may be unreliable, creating conservative or anti-conservative confidence interval coverage depending on the situation \cite{leyrat_cluster_2018}. For an overview and comparison of methods for CRT analysis, we refer the reader to \cite{hayes_cluster_2017} and \cite{benitez_defining_2022}.

Here, we address these challenges by combining \textit{Two-Stage targeted minimum loss-based estimation} (TMLE) to account for sub-sampling and missing individual-level outcomes \cite{balzer_two-stage_2021} with carefully considered \textit{conditional independence assumptions}  to address limited numbers of clusters \cite{laan_estimating_2013}. The novel contributions of this work include the following. First, we extend Two-Stage TMLE to handle differential measurement of an outcome among a closed cohort, where cohort membership is defined by sub-sampling and also subject to differential measurement at baseline. Second, we detail the assumptions required to increase the effective sample size by considering a sub-unit of the cluster to be the conditionally independent unit; this process results in the CRT behaving  like an observational study. As a consequence, we extend the prior asymptotic results and practical implementation of Two-Stage TMLE for this psuedo-observational setting. Additionally, we discuss how our approach relates to assumptions commonly made in multi-level observational studies, where, for example, individuals are nested in neighborhoods and substantial interactions occur within and across those neighborhoods \cite{Oakes2004, Sobel2006}. Finally, we demonstrate the real-life consequences of various analytic choices, using real-world data from the SEARCH-TB study.

Briefly, SEARCH-TB sought to evaluate the population-level effect of universal HIV test-and-treat on incident tuberculosis (TB) infection in rural Uganda. SEARCH-TB was a sub-study of the SEARCH trial, a 32-community CRT (NCT01864603) \cite{havlir_hiv_2019}. Intervention communities received annual, population-based HIV testing with universal treatment eligibility and patient-centered care delivery. Control communities received population-based testing at baseline with treatment eligibility according to Ministry of Health guidelines. Given logistical and financial constraints, detailed below, assessment of incident TB infection was limited to 9 communities, within which a sub-sample of participants was selected based on the HIV status of their household. Multiple visits were made to selected households to administer sociodemographic surveys and tuberculin skin tests (TSTs) to persons aged 5 years and older. The sub-study participants who were TST-negative at baseline formed a closed cohort, on whom follow-up TSTs were attempted one year later. The primary outcome of the sub-study was the one-year incidence of TB infection. The applied results have been previously presented \cite{marquez_impact_2022}; here we focus on the causal and statistical methods to account for purposefully differential sampling, potentially differential outcome measurement, and few independent units. Full discussion of the application is given in Section~\ref{results}; we now present our analytic approach more generally.


\section{Two-Stage TMLE for sampling and missing outcomes}
\label{sec:Two-stage}

In CRTs, ``two-stage'' approaches first estimate a cluster-level endpoint and then use those estimates to evaluate the intervention effect \cite{hayes_cluster_2017, murray_design_2018}. As detailed in \cite{benitez_defining_2022}, such approaches can be combined with weighting schemes to estimate cluster-level or individual-level effects on any scale. In particular, Two-Stage TMLE was developed to reduce bias and improve efficiency of CRTs by optimally adjusting for baseline cluster-level covariates, \emph{after} controlling for missingness on individual-level outcomes \cite{balzer_two-stage_2021}. In Stage 1, we identify and estimate a cluster-level endpoint, accounting for potentially differential measurement of individual-level outcomes.
To do so, we (i) define a cluster-level counterfactual parameter as a summary of the individual-level counterfactual outcomes of the cluster members, (ii) assess  identifiability of that causal parameter, and then (iii) estimate the corresponding statistical parameter in each cluster separately. In Stage 2, we use the resulting endpoint estimates from each cluster to evaluate the intervention effect, optimally adjusting for cluster-level covariates to increase precision. Two-Stage TMLE compares favorably to competing CRT methods, especially when there are post-baseline causes of missingness \cite{balzer_two-stage_2021}. We now extend the approach to account for sub-sampling and missingness at both baseline and follow-up. In Section~\ref{Sec:Est_partition}, we further extend the method to support conditional independence assumptions commonly made in  observational epidemiology.

\subsection{Stage 1: Identifying and estimating the cluster-level endpoint}
\label{stage1}

When the individual-level outcomes are not MCAR, estimating the cluster-specific endpoint with the simple mean among those measured can create several hazards. First, failing to account for over-sampling of certain subgroups and under-sampling of others can bias estimates for the population of interest. Second, in longitudinal studies, failing to account for incomplete measurement of baseline status can skew estimates of baseline prevalence and estimates of intervention effectiveness.  As an extreme example, suppose  only participants at very low risk of the outcome were tested at baseline; then estimates of baseline prevalence would be biased downwards, and the resulting incidence cohort would be a poor representation of the  population at risk. Likewise, failing to account for incomplete measurement of final endpoint status among the longitudinal cohort can also bias estimates of incidence and intervention effectiveness. As another extreme example, suppose all high-risk cohort members did not have their endpoint measured; then cluster-level estimates of incidence would be biased downwards. If missingness is present at both baseline and follow-up, these biases could compound. Further, if missingness is differential by arm --- say, the high-risk participants were more likely to be measured at follow-up in the intervention arm --- the potential for bias is even greater.

In SEARCH-TB, our motivating study, all of these dangers were present. The sub-sample was enriched for persons with HIV; measurement of baseline TB status was potentially differential among those sampled, and measurement of incident TB infection was also potentially differential among participants who were TST-negative at baseline. In the following subsection, we discuss our definition of the cluster-level endpoint and describe methods for estimating it, along with relevant assumptions.

\subsubsection{Notation.}

Throughout, we denote cluster-level quantities with superscript $c$ and underlying (possibly unmeasured) quantities with an asterisk. For an individual in a given cluster, let $E^c$ represent the cluster-level covariates (e.g., baseline HIV prevalence) and  $L_0$ the set of individual-level covariates (e.g., age). These are either measured prior to intervention implementation or, at minimum, not impacted by the intervention. Let $A^c$ represent whether the cluster was randomized to the intervention ($A^c=1$) or the control ($A^c = 0$), and $S$ indicate that an individual was sampled for the sub-study. Next, define $Y_0^*\in \{0,1\}$ as a participant's underlying (possibly unmeasured) outcome status at baseline ---  specifically, $Y_0^*=1$ if the participant has the outcome (e.g., TB infection) at baseline and 0 if not. Likewise, define $\Delta_0$ as an indicator that their outcome was measured at baseline; hence, $\Delta_0$ is deterministically 0 if the participant was not sampled ($S=0$) for the sub-study. The observed outcome at baseline is defined as $Y_0 = \Delta_0 \times Y^*_0$,  equaling 1 if the participant was measured and had the outcome at baseline. Participants known to be at risk at baseline (i.e., those with $\Delta_0=1$ and $Y_0=0$) form a closed cohort for incidence measurement. Variables $Y^*_1$, $\Delta_1$, and $Y_1$ are the follow-up timepoint analogues. Likewise, let $L_1$ denote post-baseline variables that may be impacted by the intervention $A^c$ and impact the underlying outcome $Y^*_1$  and  its measurement $\Delta_1$ at follow-up. 

\begin{figure}[!h]
\begin{center}
\includegraphics[width=\textwidth]{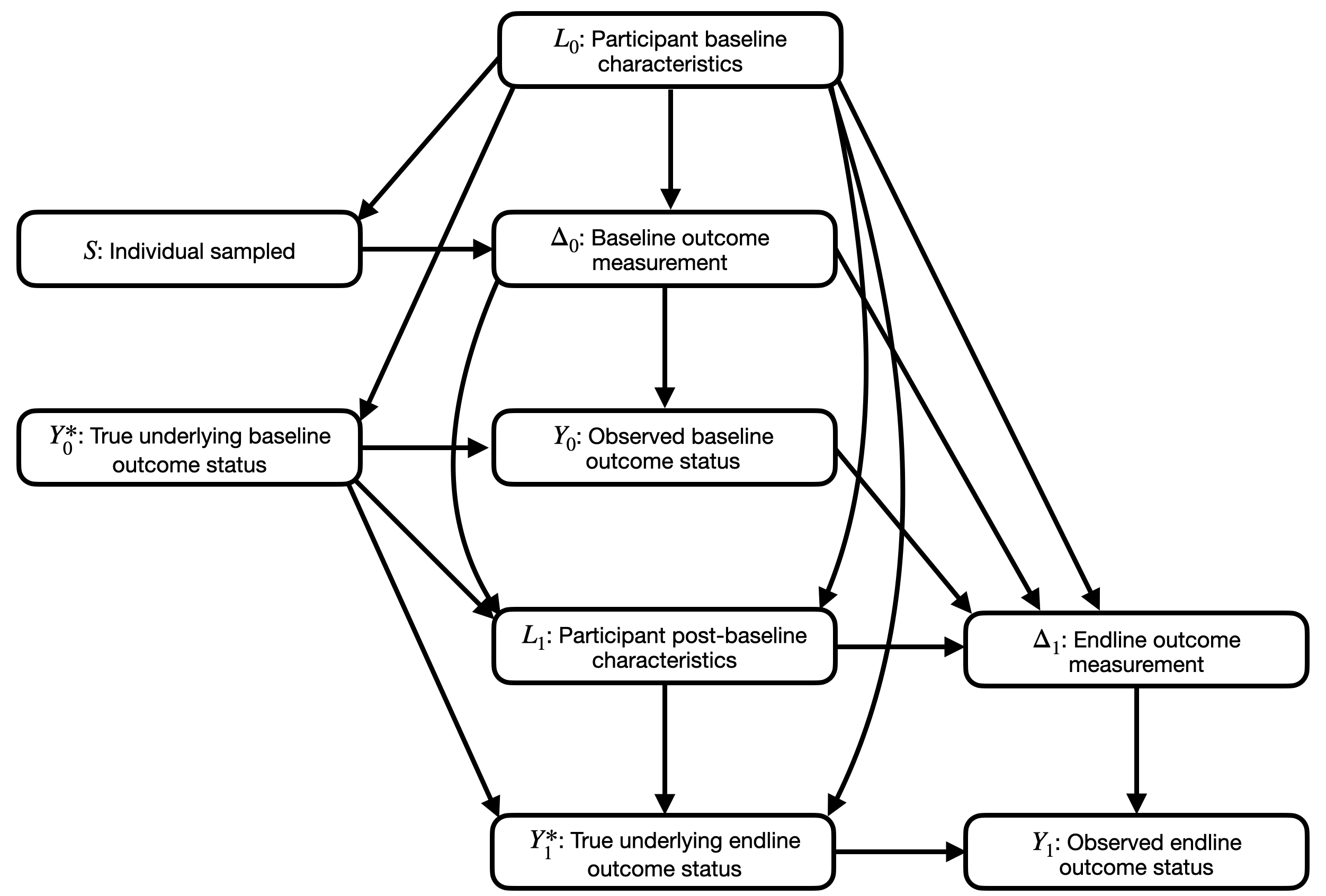}
    \caption{A simplified causal model to illustrate the relationships between individual-level variables within a cluster in Stage 1. For ease of presentation, the graph is shown without any dependence between unmeasured variables, which are omitted.}
    \label{oneclustdag}
\end{center}
\end{figure}

Altogether, the observed data for a participant are $O = (E^c,  L_0, A^c, S, \Delta_0, Y_0, L_1, \Delta_1, Y_1)$.  Recall that Stage 1 of our approach involves defining and estimating an endpoint in each cluster separately. Therefore, we can simplify the participant-level data to $O = (L_0, S, \Delta_0, Y_0, L_1, \Delta_1, Y_1)$, because the cluster-level covariates $E^c$ and cluster-level exposure $A^c$ are shared by all members of a given cluster \cite{balzer_two-stage_2021}. A simplified directed acyclic graph showing the relationships between the individual-level variables is shown in Figure \ref{oneclustdag}.

\subsubsection{Definition and identification of the cluster-level causal parameter.}
\label{sec:stage1}

In Stage 1, we focus on the underlying  proportion of cluster members with the outcome at follow-up among those at risk at baseline:
\begin{equation}
\label{eq:targparam}
\begin{aligned}
    \mathbb{P}(Y_1^* = 1 \mid Y_0^* = 0) =
    \frac{\mathbb{P}(Y_1^* = 1, Y_0^* = 0)}{\mathbb{P}(Y_0^* = 0)} = 
    \frac{\mathbb{P}(Y_1^* = 1, Y_0^* = 0)}{1-\mathbb{P}(Y_0^* = 1)}
\end{aligned}
\end{equation}
This is equivalent to the counterfactual incidence of the outcome  under the following hypothetical interventions. First, to ensure  outcome ascertainment at baseline, we would include all cluster members in the study (i.e., `set' $S=1$) and measure all participants' outcomes (i.e., `set' $\Delta_0=1$). Second, to ensure follow-up outcome ascertainment among members of the incidence cohort, we consider a  dynamic intervention to `set' $\Delta_1=1$ among those at risk at baseline $(Y_0=0, \Delta_0=1)$ \cite{hernan_comparison_2006, laan_causal_2007, robins_estimation_2008}.
We now briefly discuss the assumptions needed to express this causal parameter (Eq.~\ref{eq:targparam}) as a statistical parameter (i.e., function) of the observed data distribution. The plausibility of the identification assumptions in our motivating example is discussed in Section \ref{results} and further details are in the Supplemental Materials. 

For ease of presentation, we re-parameterize the denominator of Eq.~\ref{eq:targparam} as one minus the counterfactual outcome prevalence at baseline:  $\mathbb{P}(Y_0^*=0) = 1 - \mathbb{P}(Y_0^*=1)$. 
Under the following assumptions, the latter is identified as the baseline prevalence of the observed outcome, adjusted for differences between participants with measured versus missing outcomes: $\psi_{den}^c \equiv \mathbb{E}\{ \mathbb{E}(Y_0 \mid \Delta_0 = 1,S = 1, L_0) \}$, where superscript $c$ is used to emphasize this statistical parameter is shared by all cluster members.
To establish equivalence between $\mathbb{P}(Y_0^*=1)$ and $\psi_{den}^c$, we need that sub-sampling is done randomly within values of $L_0$ \emph{and} that the only common causes of the outcome and its measurement (among those sampled) are also captured in $L_0$. This equivalent to assuming baseline outcome status is missing-at-random (MAR): $Y^*_0 \perp \!\!\! \perp S \mid L_0$ and $Y_0^* \perp \!\!\! \perp \Delta_0 \mid S=1, L_0$. Additionally, we need a positivity assumption; sub-sampling and baseline measurement (among those sampled) is possible, regardless of $L_0$ values: $\mathbb{P}(S=1 \mid L_0 = l_0) > 0$ and $\mathbb{P}(\Delta_0 = 1 \mid S = 1, L_0 = l_0) > 0$ for all possible values $l_0 \in L_0$.

Identification of the counterfactual proportion of cluster members who have the outcome at follow-up and are at risk at baseline  $\mathbb{P}(Y_1^* = 1, Y_0^* = 0)$ is also possible under two common assumptions. First, the sequential randomization assumption \cite{robins_new_1986} requires that at each timepoint, the MAR assumption holds conditionally on (a subset of) the measured past. This assumption can be evaluated graphically with the sequential backdoor criterion \cite{pearl_causality_2009}. Second, the corresponding positivity assumption requires a positive probability of measurement at each timepoint, regardless covariate history. Under these assumptions, the numerator of the causal parameter (Eq.~\ref{eq:targparam}) can be identified as $\psi_{num}^c \equiv \mathbb{E} \left[ \mathbb{E} \{ \mathbb{E} (Y_1 \mid \Delta_1 = 1, L_1, Y_0=0,\Delta_0 = 1, S = 1, L_0) \mid \Delta_0 = 1, S = 1, L_0\} \right]$. This is the longitudinal G-computational formula expressed in terms of iterated conditional expectations \cite{Bang&Robins05, vanderLaan2012towertmle}.

Altogether, our Stage 1 statistical parameter is given by
\begin{equation}
\label{eq:stage1}
\begin{aligned}
    Y^c \equiv  \frac{\psi_{num}^c}{1-\psi_{den}^c} = \frac{ \mathbb{E} \left[ \mathbb{E} \{ \mathbb{E} (Y_1 \mid \Delta_1 = 1, L_1, Y_0=0,\Delta_0 = 1, S = 1, L_0) \mid \Delta_0 = 1, S = 1, L_0\} \right]}{1- \mathbb{E} \{ \mathbb{E}(Y_0 \mid \Delta_0 = 1,S = 1, L_0 )\} }
\end{aligned}
\end{equation}
We use $Y^c$ to emphasize this parameter is shared by all cluster members and is a summary measure of the individual-level data within that cluster. Here, $Y^c$ is a complex summary function, but can, nevertheless, be interpreted as the incidence of the outcome, after adjusting for sampling and differential measurement at both baseline and follow-up. Under the  identification assumptions, $Y^c$ would equal the counterfactual outcome incidence if there were complete sampling and no missingness: $\mathbb{P}(Y_1^*=1 | Y_0^*=0)$.

\subsubsection{Estimating the cluster-level statistical parameter.}
\label{est}

Several options exist to estimate the statistical parameters of the denominator $\psi_{den}^c$, numerator $\psi_{num}^c$, and, thus, the cluster-level endpoint $Y^c$. All approaches are implemented in each cluster separately, allowing the relationships between the individual-level covariates, sampling, measurement, and outcomes to vary by cluster \emph{and} naturally accounting for cluster-level variables $(E^c, A^c)$ \cite{balzer_two-stage_2021}. If the adjustment variables $(L_0, L_1)$ are discrete and low-dimensional,  we could implement a non-parametric stratification-based approach to estimate the iterated conditional expectations in G-computation or measurement mechanism in inverse probability weighting \cite{horvitz_generalization_1952,robins_new_1986,Bang&Robins05}.

However, when the adjustment variables are continuous and/or moderate-to-high dimensional, machine learning can be applied to avoid unsubstantiated modeling assumptions and flexibly learn complex relationships in the data. To support valid statistical inference, machine learning should be incorporated in doubly robust estimators (a.k.a, double/debiased machine learning methods), such as TMLE \cite{laan_targeted_2011, Diaz2019}. With respect to our missing data problem (i.e, estimation of $Y^c$), doubly robust estimators enjoy the following properties: asymptotic linearity under reasonable regularity conditions; consistency if either the iterated conditional expectations or the measurement mechanism is consistently estimated, and efficiency if both are consistently estimated at fast enough rates. As a substitution estimator, TMLE is often preferable to other approaches, especially under data sparsity due to positivity violations or rare outcomes. Implementation of TMLE will vary by parameter and is detailed in the Supplemental Materials for both $\psi^c_{num}$ and $\psi^c_{den}$. We recommend implementing TMLE with Super Learner \cite{laan_super_2007}, an ensemble machine learning method, to improve our chances of having both a consistent and efficient estimator.

We  obtain a point estimate of the endpoint in each cluster as $\hat{Y}^c=\hat{\psi}^c_{num}/(1-\hat{\psi}^c_{den})$. Then these estimated cluster-level endpoints $\hat{Y}^c_i$ for $i=\{1,\ldots, N\}$ are used to evaluate the intervention effect in Stage 2.

\subsection{Stage 2: Definition, estimation, and inference for the treatment effect}
\label{stage2}

Recall our goal of evaluating the intervention effect in a CRT. Let $Y^c(a^c) = \mathbb{P}(Y_1^*(a^c)=1 \mid Y_0^*(a^c)=0)$ denote the counterfactual outcome incidence under an additional intervention to `set' $A^c=a^c$. Since the cluster-level treatment is randomized, the following identification conditions hold by design: $Y^c(a^c) \perp \!\!\! \perp A^c$ and $0<\mathbb{P}(A^c=1)<1$.
Additionally, since we have already dealt with individual-level sampling and missingness in Stage 1, we can trivially identify summaries of these cluster-level counterfactuals \cite{balzer_two-stage_2021}. 
Suppose, for example, we are interested in the effect for a population of clusters; then the expected counterfactual outcome $\mathbb{E}[Y^c(a^c)]$ equals the expected outcome among those receiving the exposure of interest $\mathbb{E}(Y^c|A^c)$. Our approach  easily accommodates other effects, such conditional or sample effects; however, we focus on population effects throughout this manuscript for demonstration.

To gain efficiency, we incorporate covariate adjustment. Let $\phi^c(a^c) =\mathbb{E}\{\mathbb{E}(Y^c | A^c=a^c, E^c)\}$, where $E^c$ are the baseline covariates, including those measured directly at the cluster-level (e.g., urban vs. rural) and/or aggregates of individual-level covariates $L_0$ (e.g., HIV prevalence). $\phi^c(a^c)$ is a cluster-level analog of the G-computation identifiability result \cite{robins_new_1986, balzer_targeted_2016, balzer_new_2019}. If the Stage 1 identifiability assumptions hold, contrasts of $\phi^c(1)$ and $\phi^c(0)$ can be interpreted as the population-level intervention effects. If not, contrasts of $\phi^c(1)$ and $\phi^c(0)$ are interpreted statistically as associations of the cluster-level intervention with the incidence of the outcome, after controlling for sub-sampling and missingness at the individual-level. 

We now consider how to optimally estimate the Stage 2 statistical parameter $\phi^c$, defined as a contrast between  $\phi^c(1)$ and $\phi^c(0)$. For example, on the relative scale, $\phi^c = \phi^c(1) \div \phi^c(0)$. In Stage 2, our observed  data are at the cluster level: $O^c = (E^c, A^c, \hat{Y}^c)$, where $\hat{Y}^c$ is the cluster-level endpoint estimated in Stage 1.

Using these data, Stage 2 estimation can proceed by implementing a cluster-level analysis, such a G-computation, inverse probability weighting, or TMLE. The key challenge to Stage 2 is \emph{a priori} specification of the optimal adjustment set --- which variables and what functional form.  One solution to this challenge is to implement \textit{Adaptive Pre-specification} (APS) within TMLE \cite{balzer_adaptive_2016}. Briefly, APS pre-specifies a candidate set of working generalized linear models (GLMs) for the cluster-level outcome regression $\mathbb{E}(\hat{Y}^c | A^c, E^c)$ and for the cluster-level propensity score  $\mathbb{P}(A^c=1| E^c)$ and, then, chooses the combination that minimizes the cross-validated variance estimate for the TMLE of the target parameter. Finite sample simulations and real-data applications  have demonstrated substantial precision gains over alternative approaches \cite{balzer_adaptive_2016, balzer_two-stage_2021, benitez_defining_2022}.

Under conditions detailed in \cite{balzer_two-stage_2021}, the Two-Stage TMLE $\hat{\phi}^c$ will be an asymptotically linear estimator of the target effect $\phi^c$, such that $\hat{\phi}^c - \phi^c = 1/N \sum_{i=1}^N D^c_i + R_N$ with $D^c_i$ as the influence curve (function) for the $i^{th}$ cluster and $R_N=o_p(1/\sqrt{N})$ as the remainder term \cite{vanderVaart1998}. In particular, we need the contributions from Stage 1 estimation to the remainder term $R_N$ to be essentially zero. Practically, this means we should not bet on bias cancellations when defining or estimating the  cluster-level endpoint $Y^c = \psi^c_{num}/(1-\psi^c_{den})$. Indeed, biased estimators of the cluster-level endpoints can result in biased estimates of and misleading inference for the intervention effect. Instead, we recommend using TMLE, incorporating machine learning, to flexibly estimate the cluster-level endpoint $Y^c_i$ for $i=\{1,\ldots, N\}$ in Stage 1. Additionally, two-stage approaches are most effective when the cluster size is relatively large, allowing for adaptive and well-supported estimation of the cluster-level endpoints. The  regularity conditions required of Stage 2 estimators of the cluster-level outcome regression and known propensity score hold by design, when using APS to select from working GLMs in TMLE.  As discussed next, however, the conditions on the Stage 2 estimators will change if we make alternative  identification assumptions in Stage 2.

Under the above conditions, Two-Stage TMLE will be normally distributed in the large data limit, allowing for the construction of Wald-type confidence intervals as 
$\hat{\phi}^c \pm 1.96 \hat{\sigma}$, where $\hat{\sigma}^2$ is the sample variance of the estimated cluster-level influence curve $\hat{D}^c$, scaled by sample size $N$. (The form of the influence curve will depend on the target parameter $\phi^c$.) In CRTs with fewer than 40 clusters randomized ($N<40$), we recommend using the Student's $t$ distribution with $N-2$ degrees of freedom as a finite sample approximation of the asymptotic normal distribution \cite{hayes_cluster_2017}.

\section{(Re-)Defining the independent unit}
\label{2comm}

A fundamental premise of CRTs is that outcomes  are dependent within a cluster. Sources of dependence could include shared cluster-level factors, including the intervention, as well as social  interactions between participants within a cluster. Instead, clusters are assumed to be independent, providing the basis for statistical inference, as described in the prior subsection. However, CRTs tend to randomize few clusters, limiting  statistical power. For example, while its parent trial randomized 32 communities, measurement of incident TB infection in  SEARCH-TB occurred in only 9 communities in  Uganda. Even if a given CRT has many clusters, subgroup analyses to understand effect heterogeneity may be conducted among limited numbers of clusters. The extreme case of randomizing to  only two clusters, a \emph{de facto} observational study, was covered in depth by \cite{laan_estimating_2013}.

In this section, our goals are to (i) define a hierarchical causal model, reflecting the data generating process for a CRT, (ii) detail the assumptions needed to consider a sub-unit of the cluster to be the conditionally independent unit, and (iii) present the consequences of these assumptions for statistical estimation and inference with Two-Stage TMLE. The level of clustering and, thereby, the definition of ``sub-unit'' will vary by setting. In SEARCH-TB, for example, individuals are nested within households, villages, parishes, and  communities. Under different assumptions, explicitly stated below, any level of partitioning of the cluster could be treated as the conditionally independent unit. 

For simplicity, we focus on CRTs with 3 layers of clustering: individuals are grouped into sub-cluster ``partitions", indexed  by $j = \{1, \ldots,J\}$, and these partitions are grouped into a cluster, which remain the unit of randomization. As before, we denote cluster-level variables with superscript $c$. Now, denote partition-level variables with superscript $p$. Recall $E^c$ is the set of cluster-level characteristics; these are sometimes called ``environmental'' factors, because they represent the shared environment of individuals in a given cluster \cite{laan_estimating_2013}. As before, $A^c$ is an indicator of the cluster being  randomized to the intervention arm. Now, let $W_j^p$ be the set of baseline covariates for partition $j$; these could be general characteristics of the partition (e.g., urban vs. rural) as well as aggregates of baseline covariates of individuals from that partition (e.g., HIV prevalence). Likewise, let $Y_j^p$ be the $j^{th}$ partition's endpoint, which is defined analogously to $Y^c$ in Stage 1 (Eq.~\ref{eq:stage1}). Specifically, $Y_j^p$ is the incidence of the outcome, after adjusting for sampling and differential measurement among members of partition $j$. Under the identification assumptions given in Section~\ref{sec:stage1}, $Y_j^p$ would equal the counterfactual incidence of the outcome for partition $j$ if we had complete sampling and no missingness.

\subsection{Hierarchical structural causal models}
\label{ex0}

Using the non-parametric structural causal model of \cite{pearl_causality_2009}, we now formalize the hierarchical data generating process for a CRT. For ease of presentation, we  focus on CRTs with $J=2$ partitions per cluster; however, our results naturally generalize to other settings.

\begin{figure}[!h]
\begin{center}
\includegraphics[width=\textwidth]{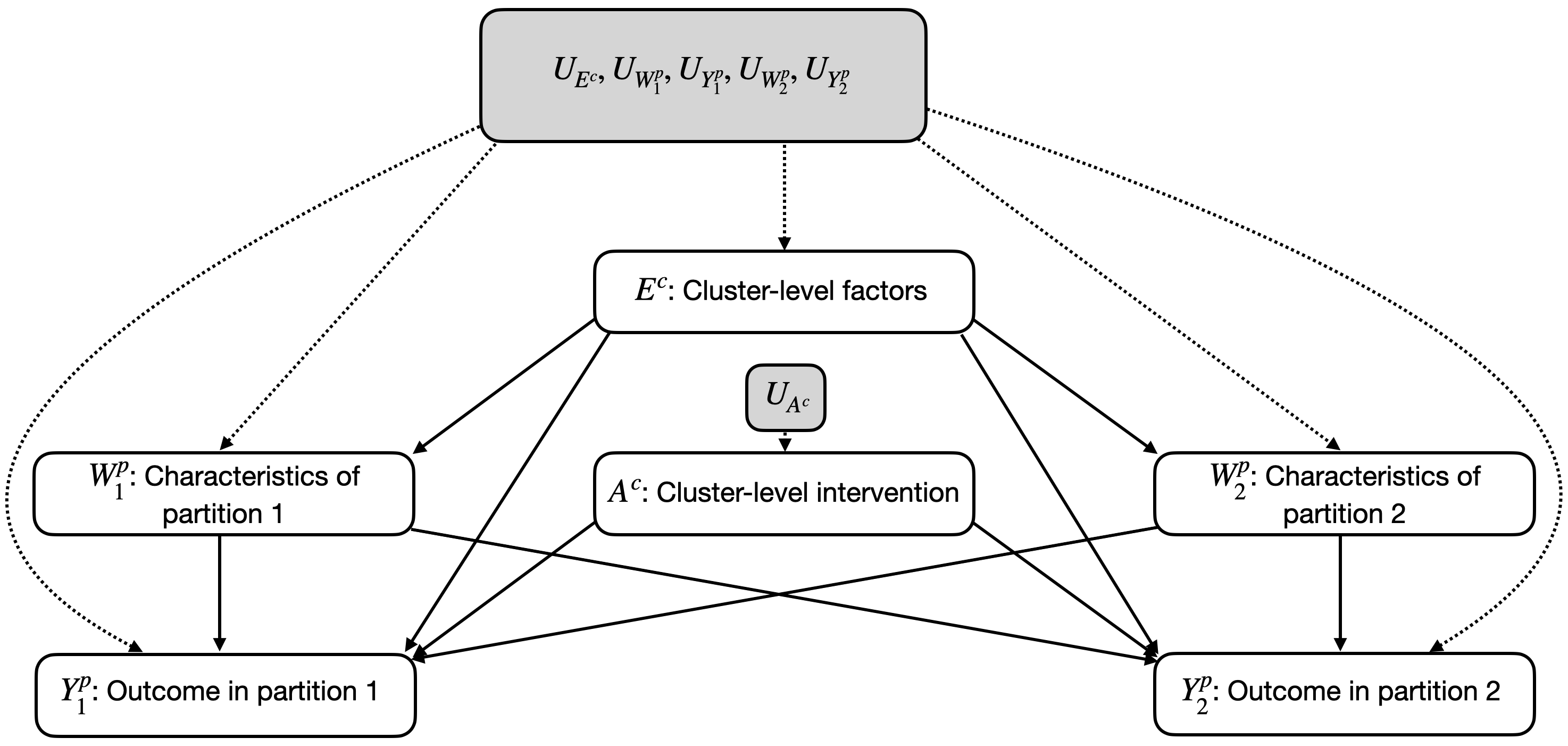}
    \caption{A simplified causal model for the data generating process of a cluster randomized trial with 2 partitions (i.e., sub-units) per cluster. By design in a cluster randomized trial, the unmeasured factors contributing to the cluster-level intervention $U_{A^c}$ are independent of the others. We make no other exclusion restrictions or independence assumptions.}
    \label{s2_u}
\end{center}
\end{figure}

Figure~\ref{s2_u} provides a causal model, assuming independence between clusters and randomization of the cluster-level intervention $(U_{A^c} {\perp\!\!\!\perp} U_{E^c}, U_{W_1^p}, U_{W_2^p}, U_{Y_1^p}, U_{Y_2^p})$. The structure of the remaining $U$s may be complex and cluster-specific; for example, the unobserved factors influencing the partition-level outcomes $(U_{Y^p_1}, U_{Y^p_2})$ might be related to unmeasured, environmental factors $U_{E^c}$. Beyond the unmeasured factors, there are several sources of dependence between partition-level outcomes in this model. For example, the $j^{th}$ partition's outcome $Y^p_j$ may depend on the characteristics of the other $W^p_{-j}$. This general causal model encodes independence at the cluster-level, not the partition-level --- yet.

\begin{figure}[!ht]
\begin{center}
\includegraphics[width=\textwidth]{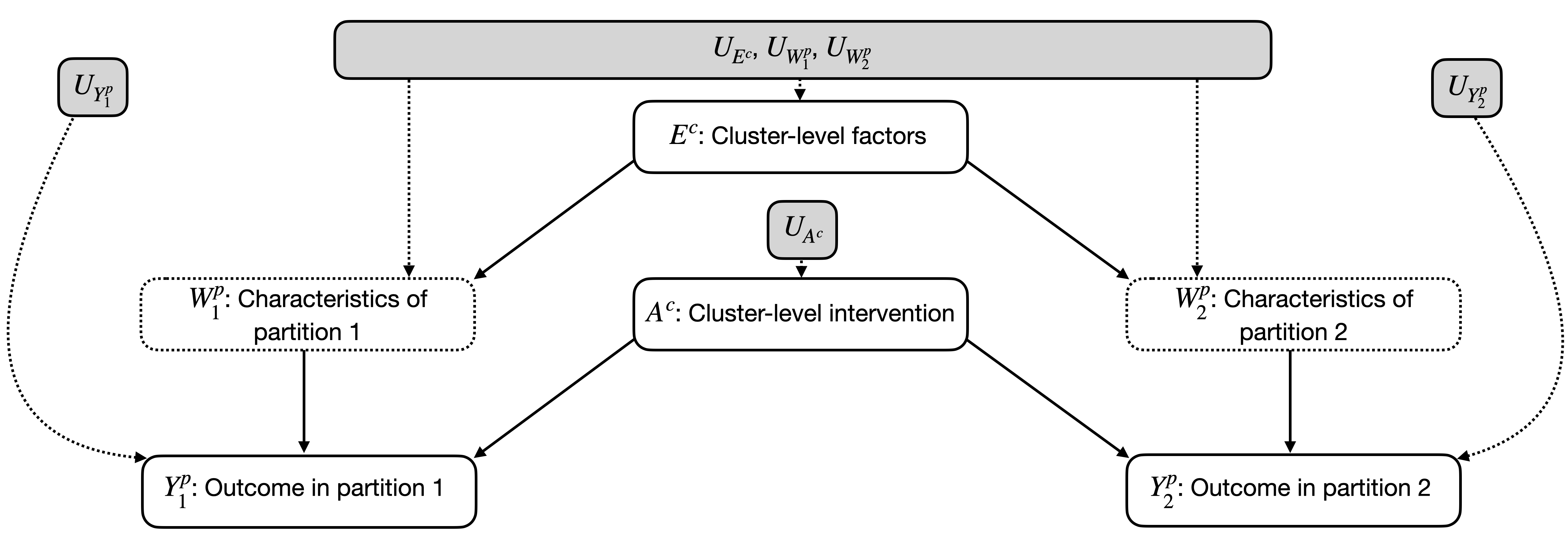}
    \caption{A restricted causal model for the data generating process of a cluster randomized trial with 2 partitions (i.e., sub-units) per cluster and under the assumptions needed for the partitions to be conditionally independent. This graph reflects the following exclusion restrictions and independence assumptions: no interference between partitions; no direct effect of the cluster-level covariates $E^c$ on the partition-level outcomes $Y^p$, and no unmeasured common cause of the partition-level outcomes $(U_{Y^p})$ and the cluster-level or partition-level covariates ($U_{E^c},U_{W^p}$). By design in a cluster randomized trial, the unmeasured factors contributing to the cluster-level intervention $(U_{A^c})$ are independent of the others.}
    \label{s2_indep}
\end{center}
\end{figure}

To treat the sub-cluster partition as the conditionally independent unit, we need several assumptions to hold, resulting in a more restrictive causal model reflected in Figure~\ref{s2_indep} \cite{laan_estimating_2013}.
First, there is no interference between partitions within a cluster. Second, any effect of the cluster-level covariates $E^c$ on the partition-level outcome $Y_j^p$ is only through their effect on $j^{th}$ partition's covariates $W^p_j$. Finally, there are no unmeasured common causes of partition-level outcomes $Y_j^p$ and the cluster-level or partition-level covariates ($E^c, W^p_j)$. While we additionally need the unmeasured factors contributing to the cluster-level intervention $A^c$ to be independent of the others, this holds by design in CRT. Altogether, these assumptions require there to be no interactions between partitions within a cluster \emph{and} the partition-level covariates $W^p$ are sufficient to block the effects of the cluster-level, environmental factors $E^c$ on the partition-level outcomes $Y^p$. If these assumptions hold, the partition becomes the conditionally independent unit, increasing the effective sample size, while still allowing for arbitrary dependence \emph{within} each partition.

Whether or not these assumptions are reasonable depends on the study context. To maximize the effective sample size, it might be tempting to define the ``partitions" as the individuals in a cluster. However, this approach would entail very strong and possibly unrealistic assumptions, especially  in the setting of infectious or contagious outcomes. Instead, if partitions are large sub-units of the cluster (e.g., distant neighborhoods in a rural community), these assumptions might be reasonable. Altogether, the assumptions needed to treat the partition as the conditionally independent unit are strong; however, they are commonly evoked in multi-level, observational epidemiology \cite{Oakes2004, Sobel2006}. By explicitly stating them and illustrating them with a causal graph, we aim to empower readers to judge whether they are plausible. Additionally, the design of future studies can be improved by measuring a rich set of covariates to improve the plausibility of these assumptions.

\subsection{Estimation and inference with partition-level conditional independence}
\label{Sec:Est_partition}

The assumptions encoded in the restrictive causal model (Figure~\ref{s2_indep}) have important implications for our two-stage estimation approach. Previously, when considering the cluster to be the independent unit, we identified and estimated a cluster-level endpoint $Y^c$ that accounted for sub-sampling of individuals within that cluster, missingness on baseline outcome status of sampled individuals, and missingness on final outcome status of individuals known to be at risk at baseline. Under the more restrictive model, we now identify and estimate a partition-level endpoint $Y^p$ in Stage 1. Practically, this means that within each partition separately, we use TMLE to estimate $Y^p=\psi^p_{num}/(1-\psi^p_{den})$, as defined  in Eq.~\ref{eq:stage1}, and then use the resulting estimates $\hat{Y}^p$ to evaluate the intervention effect in Stage 2.

During effect estimation in Stage 2, we previously adjusted for cluster-level covariates $E^c$ simply to increase precision in a CRT. Now, however, blurring the lines between randomized trials and observational studies \textit{requires} us to adjust for confounders $W^p$ to identify the causal effect and support the conditional independence assumptions. Recall adjustment for the partition-level covariates $W^p$ is required to block the effect of the cluster-level environmental factors $E^c$, which are no longer included in the adjustment set. Therefore, the Stage 2 statistical estimand is now defined in terms of contrasts of the expected partition-level endpoint, given the cluster-level treatment and partition-level confounders: $\phi^p(a^c) =  \mathbb{E}\{\mathbb{E}(Y^p | A^c=a^c, W^p)\}$. For example, on the relative scale, our statistical estimand would be  $\phi^p =  \phi^p(1) \div \phi^p(0)$. As noted earlier, our approach can target other effects, such as the conditional or sample effects, defined on any scale.)

Importantly, the revised statistical estimand $\phi^p$  has a subtly different interpretation than the original statistical estimand $\phi^c$, which was in terms of the expected cluster-level outcome. If the number of partitions per cluster varies, the value of these two estimands could differ; however, weights can be applied to recover either estimand \cite{benitez_defining_2022}. Statistically, $\phi^p$ can be interpreted  as the association of the cluster-level intervention with the incidence of the outcome, after controlling for sub-sampling and missingness at the individual-level and for confounding from environmental factors at the partition-level. However, if the Stage 1 identifiability assumptions hold \emph{and} the Stage 2 identifiability assumptions hold,  $\phi^p$ can be interpreted as the population-level intervention effect. 
The revised Stage 2 statistical estimand $\phi^p$ could be estimated with a variety of methods. We again recommend TMLE, given its double robustness property, potential for efficiency, and  ability to incorporate machine learning while maintaining the basis for valid statistical inference. To implement TMLE for $\phi^p$ in this setting, we pool together partition-level observations $O^p_k = (W^p_k, A^c_k, \hat{Y}^p_k)$ for the $k=\{1,\ldots, K\}$ partitions in the CRT.  Now, $\hat{Y}^p$ represents the estimated partition-level endpoint from Stage 1. Using these data, we implement TMLE at the partition-level as if we had a point-treatment observational study \cite{laan_targeted_2011}.

Treating the partition as the conditionally independent unit  changes our approach to statistical inference. Specifically, our effective sample size is now $K$, the number of partitions. However, this comes at the cost of stronger conditions for Two-Stage TMLE $\hat{\phi}^p$  to be asymptotically linear for the target parameter $\phi^p$, such that $\hat{\phi}^p - \phi^p = 1/K \sum_{i=1}^K D^p_k+ R_K$ with $D^p_k$ as the influence curve (function) for the $k^{th}$ partition and $R_K=o_p(1/\sqrt{K})$ as the remainder term. Now, we need the Stage 1 estimators of the partition-level endpoint $Y^p$ to contribute negligibly to the remainder term \cite{balzer_two-stage_2021}. Furthermore, the regularity conditions on effect estimation in Stage 2 do not hold by design. Instead, we need estimators of the partition-level outcome regression $\mathbb{E}(\hat{Y}^p \mid A^c, W^p)$ and partition-level propensity score $\mathbb{P}(A^c \mid W^p)$ to converge to the truth at quick enough rates and avoid overfitting \cite{laan_targeted_2011}. To satisfy these conditions, we again recommend implementing TMLE with Super Learner, considering a diverse set of candidate algorithms, in both Stage 1 and Stage 2.

\section{Application to the SEARCH-TB Study}
\label{results}

An estimated 1.7 billion people, approximately a quarter of the world’s population, are infected with tuberculosis (TB), and this vast reservoir 
fuels TB disease and death \cite{houben_global_2016, macpherson_mortality_2009}. Understanding TB transmission dynamics and then implementing effective public health interventions is difficult \cite{marquez_impact_2022}. First, transmissions are airborne and likely occur both inside and outside the household. Second, the focus has largely been on active TB (i.e., TB disease),  missing the majority of transmission events, which are latent infections. Finally, measurement of latent TB infection through tuberculin skin tests (TSTs) is expensive and imperfect.

Due to resource constraints, evaluation of SEARCH's universal HIV test-and-treat intervention on incident TB infection was conducted through a sub-study known as SEARCH-TB  \cite{marquez_impact_2022}. This sub-study was limited to 9 communities in eastern Uganda and 100 randomly sampled households in each community. As previously discussed, househod sampling was enriched for persons with HIV.  Among members of the sampled households, latent TB infection was measured via door-to-door placement and reading of TSTs. Incident TB infection was defined as conversion from a negative to positive TST after one-year of follow-up. Finally, given few randomized clusters, \emph{parishes}, a sub-unit of the community (analagous to the partitions discussed in Section~\ref{2comm}), were considered to be the conditionally independent unit under the assumptions detailed below.

\subsection{Stage 1: Identification and estimation of the one-year incidence of TB infection in each partition} 

We first defined and estimated a partition-level endpoint $Y^p$, appropriately accounting for sub-sampling and differential TB ascertainment at the individual-level. Of the 17,858 households in the 9 study communities, 1435  were sampled, and 688 (47.9\%) of the sampled households had at least one adult (aged 15 and up) with HIV. The adult  prevalence of HIV in the sub-sample was 19.6\%, a sharp contrast to the prevalence in the region of 3.6\% \cite{havlir_hiv_2019}.  Since the risk of TB differs by HIV serostatus \cite{macpherson_mortality_2009}, ignoring the sampling scheme would bias estimates of TB burden and the intervention effect. However, sampling $S$ was random within household HIV status $H$. Thus, the following assumptions were satisfied by design: $Y^*_0 \perp \!\!\! \perp S \mid H$ and $\mathbb{P}(S=1 \mid H=h)>0$ for $h\in\{0,1\}$.

Despite up to three visits to the sampled households, including weekends and after hours, TSTs were administered to 4884/8420 (58\%) of household members at baseline. Known risk factors for prevalent TB and missingness include
age and mobility \cite{marquez_impact_2022}. Let $W$ represent these baseline individual-level risk factors. We were willing to assume that for sampled individuals and within values of $W$, TB prevalence among those with a baseline TST was representative of TB prevalence among those without a baseline TST: $Y_0^* \perp \!\!\! \perp \Delta_0 \mid W, S=1, H$.  Additionally, we assumed that among those sampled, there was a positive probability of administering a TST within all possible values of $W$. These assumptions, together with the sampling design, allowed for the identification of the counterfactual baseline TB prevalence in each partition $\mathbb{P}(Y_0^*=1)$ as $\psi_{den}^p = \mathbb{E}\{ \mathbb{E}(Y_0 \mid \Delta_0=1, W, S=1, H)\}$. 

Among the 4884 participants with known baseline TB status, 3871 (79\%) were TST-negative, forming a closed cohort for incidence measurement. As before, despite best efforts, follow-up TST administration was imperfect, with 3003/3871 (78\%) of the cohort measured at follow-up. To address potentially differential ascertainment of follow-up status, we  considered common risk factors for incident TB infection and its measurement. Given the epidemiology of the region, we again identified age, mobility, and household HIV status as key joint causes of outcomes and missingness. We  assumed that within values of these adjustment factors, the risk of incident TB infection among cohort members with a follow-up TST  was representative of the risk  among cohort members without a follow-up TST. We also assumed a positive probability of receiving a follow-up TST (among the incidence cohort) within all values of $(W,H)$. These assumptions were again supported by the study design, including the repeat visits to households, and allowed for identification of the counterfactual proportion with TB at follow-up but not at baseline $\mathbb{P}(Y_1^*=1, Y_0^*=0)$. The corresponding statistical estimand was $\psi_{num}^p = \mathbb{E}[ \mathbb{E}\{ \mathbb{E}( Y_1 \mid \Delta_1 = 1, Y_0=0 ,\Delta_0 = 1, W,  S = 1, H) \mid \Delta_0 = 1, W, S = 1, H\}]$, which is a simplified version of the $\psi^c_{num}$ parameter from Section~\ref{stage1} but without the time-dependent covariates $L_1$.

For estimation and inference in Stage 1, we stratified on parish, the assumed conditionally independent unit, and estimated $\psi^p_{num}$ and $\psi^p_{den}$ with a participant-level TMLE using Super Learner to combine predictions from main-terms GLM, multivariate adaptive regression splines, and the simple mean. Then for each parish,  we obtained estimates of the one-year incidence of TB infection as $\hat{Y}^p=\hat{\psi}^p_{num} \div (1-\hat{\psi}^p_{den})$.

\subsection{Stage 2: Evaluation of the intervention effect in SEARCH-TB}

Next, we used the Stage 1 endpoint estimates $\hat{\psi}^p_k$ for  $k=\{1,\ldots, K\}$ to evaluate the effect of the cluster-level intervention in Stage 2. Before doing so, we needed to critically evaluate the assumptions needed to treat the $K$ sub-community partitions as the conditionally independent unit. Given the following considerations, we immediately eliminated the individual and the household as possible candidates. First, factors influencing TB infection risk include both an individual's susceptibility (e.g., age and HIV status) as well as their level of exposure to TB. Within a household, one member's risk factors could influence their own TB status as well as the TB status of the other household members, especially in settings with poor ventilation and shared sleeping areas. This directly violates the assumption of no interference between individuals within a household. Furthermore, an estimated 80\% of TB cases are acquired outside of the household \cite{martinez_transmission_2017, martinez_paediatric_2019} --- violating the potential assumption of no interference between households.

Therefore, for the following reasons, we assumed the parish, a large sub-unit of the community, to be the conditionally independent unit. First, we considered how and where TB is transmitted outside the home in rural Ugandan communities. Prior studies from high-TB burden countries in Sub-Saharan Africa have shown clinics, schools, churches, and workplaces are the areas of high TB risk \cite{andrews_integrating_2014}. Additionally, prior molecular epidemiologic studies in Uganda have highlighted the role of bars in TB transmission  \cite{chamie_identifying_2015, chamie_spatial_2018}. After conducting community mapping and having detailed discussion with the larger Ugandan research team, we concluded these locations are generally shared within a parish, but it was unlikely people would travel between parishes to visit these locations. Therefore, we were willing to assume that there was negligible interference between parishes within a commmunity.
 
We then considered whether the measured parish-level covariates were sufficient to block the effects of the environmental, community-level factors.  First, the role of HIV in fueling the TB epidemic is  well-established;  the biomedical mechanism is via immunosuppression leading to increased susceptibility to infection and reactivation of latent TB infections  \cite{getahun_hiv_2010}.  
Additionally, the relationship between TB and alcohol has been well-established. Globally, an estimated 10\% of TB disease is attributable to alcohol use disorder \cite{rehm_association_2009}, and a large systematic review found a 3-fold higher risk of TB disease associated with alcohol use disorder \cite{lonnroth_alcohol_2008}. Our team's prior research in Uganda has also demonstrated a dose-response relationship between levels of alcohol use and latent TB infection \cite{puryear_higher_2021}. The underlying mechanisms include alcohol-induced immunosuppression and increased exposure to TB due to time-spent in bars, which are high-TB-risk venues. Altogether, we were willing to assume that the parish-level characteristics of HIV prevalence and prevalence of adults who drink alcohol $W^p$ were sufficient to block the influences of other community-level covariates $E^c$ on the one-year incidence of TB infection in each parish $Y^p$. Under these assumptions and with 2 parishes per community, the effect sample size was $K=18$. For estimation and inference of the relative effect $\phi^p = \mathbb{E}\{\mathbb{E}(Y^p | A^c=1, W^p)\} \div  \mathbb{E}\{\mathbb{E}(Y^p | A^c=0, W^p)\} $ in Stage 2, we implemented a parish-level TMLE with Super Learner using the same library of prediction algorithms. Computing code is available at \url{https://github.com/joshua-nugent/search-tb}.

\subsection{Results of the real-data analysis}
\label{comparison}

The results of the SEARCH sub-study on incident TB infection have been previously presented in \cite{marquez_impact_2022}. The primary pre-specified analysis, using Two-Stage TMLE with the parishes as the conditionally independent unit, suggested that the universal HIV test-and-treat intervention resulted in a 27\% reduction in incident TB infection in eastern Uganda; the adjusted relative risk (aRR) was $0.73$ $(95\%$ CI: $0.57-0.92;$ $p$=0.005).

We now explore the practical impact of varying the identfication assumptions on estimation and inference. The results of our comparison are summarized in Figure~\ref{results_fig} and Table {\color{red}1} in the Supplemental material. First, we relaxed the assumption that parishes were conditionally independent and, instead, took a more traditional approach treating the randomized unit (i.e., the community) as the independent unit. As expected, when we moved from a parish-level analysis ($K=18$) to a community-level analysis ($N=9$), the effect estimate shifted  and substantial precision was lost: aRR $= 0.93$ $(95\%$ CI: $0.66-1.31;$ $p = 0.32)$. In this secondary analysis, Stage 1 was implemented analogously to obtain community-level estimates of TB incidence, accounting for sampling and missingness at the individual-level. However, Stage 2  effect estimation was done at the community-level with TMLE, using Adaptive Pre-specification to select the adjustment covariates to maximize empirical efficiency in the CRT \cite{balzer_adaptive_2016}.

To further explore the impact of our assumption that parishes were conditionally independent, we conducted a sensitivity analysis where Stage 1 accounted for missingness (as before), but Stage 2 was implemented without adjustment. This approach corresponds to the very strong assumption that the only source of dependence between parishes was the shared community-level intervention $A^c$. In other words, this analysis assumed no community-level covariates (measured or not) directly or indirectly influenced the incidence of TB infection. Estimates from this approach were again in the similar direction, but even less precise: aRR $= 0.91$ $(95\%$ CI: $0.63 - 1.32$; $p=0.30)$.

Next we explored the impact our missing data assumptions. Specifically, we conducted a sensitivity analysis where Stage 1 estimates of incidence were unadjusted, but Stage 2 was adjusted (as before). This approach corresponds to the very strong and unreasonable assumption that individual-level outcomes were missing completely at random (MCAR). In fact, we know this assumption was violated: the sub-sample was enriched for persons with HIV, and HIV is a known risk factor for TB. Age and mobility are additional risk factors for TB and for not having a TST administered at baseline or follow-up. Estimates from the approach were markedly different and in the opposite direction of the primary analysis: aRR $=1.05$ $(95\%$ CI: $0.80 - 1.37$; $p=0.63)$. In other words, conducting a complete-case analysis would lead to the conclusion that the SEARCH intervention \textit{increased} the incidence of TB infection by 5\%.

Finally and as an extreme example of strong assumptions on measurement and dependence, we conducted a fully unadjusted analysis. In Stage 1, we estimated the parish-level incidence of TB infection with the raw proportion among those measured. Then in Stage 2, we compared parish-level incidence estimates by arm without further adjustment. This approach is not recommended in practice and suggested the SEARCH intervention \textit{increased} the incidence of TB infection by 18\%: aRR $= 1.18$ $(95\%$ CI: $0.85-1.63$; $p=0.84)$.

\section{Discussion}
\label{discussion}

Cluster randomized trials (CRTs) allow for the rigorous evaluation of interventions delivered at the group-level. Within CRTs, rare or expensive outcomes may only be measured in a subset of clusters and, within those clusters, on a sub-sample of participants. Missing outcomes among participants is another common issue, which can bias estimates of baseline prevalence, the incidence of the outcome, and the intervention effect. To address these challenges, we extended Two-Stage TMLE to account for sub-sampling of participants and differential measurement of their outcomes at baseline and at follow-up. Additionally, we detailed the assumptions needed to consider a sub-cluster partition as the conditionally independent unit. We also extended Two-Stage TMLE to this novel setting, which blurs the lines between CRTs and observational studies. Our application to real-data from SEARCH-TB  demonstrated the real-world impact of varying assumptions and analytic choices. For example, ignoring the sampling scheme and assuming the outcomes were missing-completely-at-random reversed the direction of the estimated intervention effect.

When estimating the endpoint in Stage 1 and evaluating the intervention effect in Stage 2, we  used TMLE with Super Learner to avoid parametric assumptions and, instead, support efficient estimation in large, semi-parametric models. In the absence of missing data, a single-stage approach, such as GLMMs or GEE, could be used to estimate the intervention effect if the effective sample size is sufficiently large. These methods account for the dependence of participants within a partition and can incorporate adjustment for partition-level variables $W^p$ needed to support the independence assumptions. However, when adjusting for covariates, these alternative estimators are often limited in their ability to estimate marginal effects \cite{benitez_defining_2022}. For example, when using the logit-link in GLMM and GEE, the conditional odds ratio is estimated \cite{laird_random-effects_1982, hubbard_gee_2010}. Additionally, as previously discussed, even after considering the sub-cluster partition to be the conditionally independent unit, the effective sample size may still be too small to support use of these approaches without finite sample corrections. Finally and perhaps most importantly, these methods cannot accommodate post-baseline causes of missingness \cite{balzer_two-stage_2021}. Altogether, to handle common analytic challenges in CRTs (e.g., differential missingness and few clusters) and  to estimate marginal effects on any scale, we recommend using TMLE, a doubly robust, semi-parametric efficient, substitution estimator, in our two-stage approach.

Nonetheless, our approach does require real assumptions on the missngness mechanism and the dependence structure within a cluster. These assumptions have implications for trial design. First, all the shared causes of missingness and outcomes must be measured. Second, fairly large cluster sizes (or sub-cluster partition sizes) are needed for stable and consistent estimation of the endpoints in Stage 1. 
Finally, to support any conditional independence assumptions and improve precision in Stage 2, a rich set of partition-level covariates should be collected. We again emphasize these conditional independence assumptions are commonly made, but less commonly acknowledged, in multi-level observational studies \cite{Oakes2004, Sobel2006}.

In all cases, these assumptions should be carefully considered, transparently stated, and illustrated with a causal graph. As discussed in the real-data example, assuming individuals or households are effectively independent might be unrealistic in many settings. Alternatively, considering larger partitions of the cluster, such as distant neighborhoods, might be more reasonable.  While larger partitions weakens the required identification assumptions, fewer (conditionally) independent units raise finite sample concerns for estimation and inference in Stage 2. Specifically, there can arise a tension between adjusting for too many partition-level covariates (with the potential of overfitting, even with cross-validation) and including too few (not supporting the identification assumptions). In future work, we plan to use `collaborative' TMLE \cite{laan_collaborative_2010} where the  partition-level propensity score would be fit in response to adjustment conducted in the partition-level outcome regression. As illustrated with the real-data example, in-depth discussion with subject matter experts is imperative to identifying the minimal adjustment set needed to support our assumptions --- both on the missingness mechanism and on within cluster dependence. Conducting a simulation study, informed by the real-data application, can help guide development of the statistical analysis plan.

This work addresses 4 common challenges in the design and analysis of CRTs: (i)	sub-sampling of participants for measurement of a rare or expensive outcome; (ii) missingness on the baseline outcome status of sampled participants; (iii) missingness on the final outcome status of participants known to be ``at-risk'' at baseline; and (iv) very few independent units (i.e., clusters). To address the first 3 challenges, we extended Two-Stage TMLE to account for potentially biased sampling and outcome measurement. To address the final challenge, we carefully articulated and critically evaluated the assumptions required to treat sub-cluster partitions as conditionally independent. These assumptions increase our effective sample size, at the cost of making the CRT behave more like an observational study.


\section*{Acknowledgements}
\singlespacing

We extend many thanks to the Ministries of Health of Uganda and Kenya; our research and administrative teams in San Francisco, Uganda, and Kenya; our collaborators and advisory boards; and, especially, all the communities and participants involved. We also thank Drs. Lina Montoya and Mark van der Laan for their expert advice. Funding support for this work was provided by The National Institutes of Health (U01AI099959, UM1AI068636, K23AI118592, and R01AI151209) and the President's Emergency Plan for AIDS Relief.

{\it Conflict of Interest}: None declared.

\clearpage

\newpage

\section*{Tables and figures}

\begin{figure}[!h]
\begin{center}
\includegraphics[width=\textwidth]{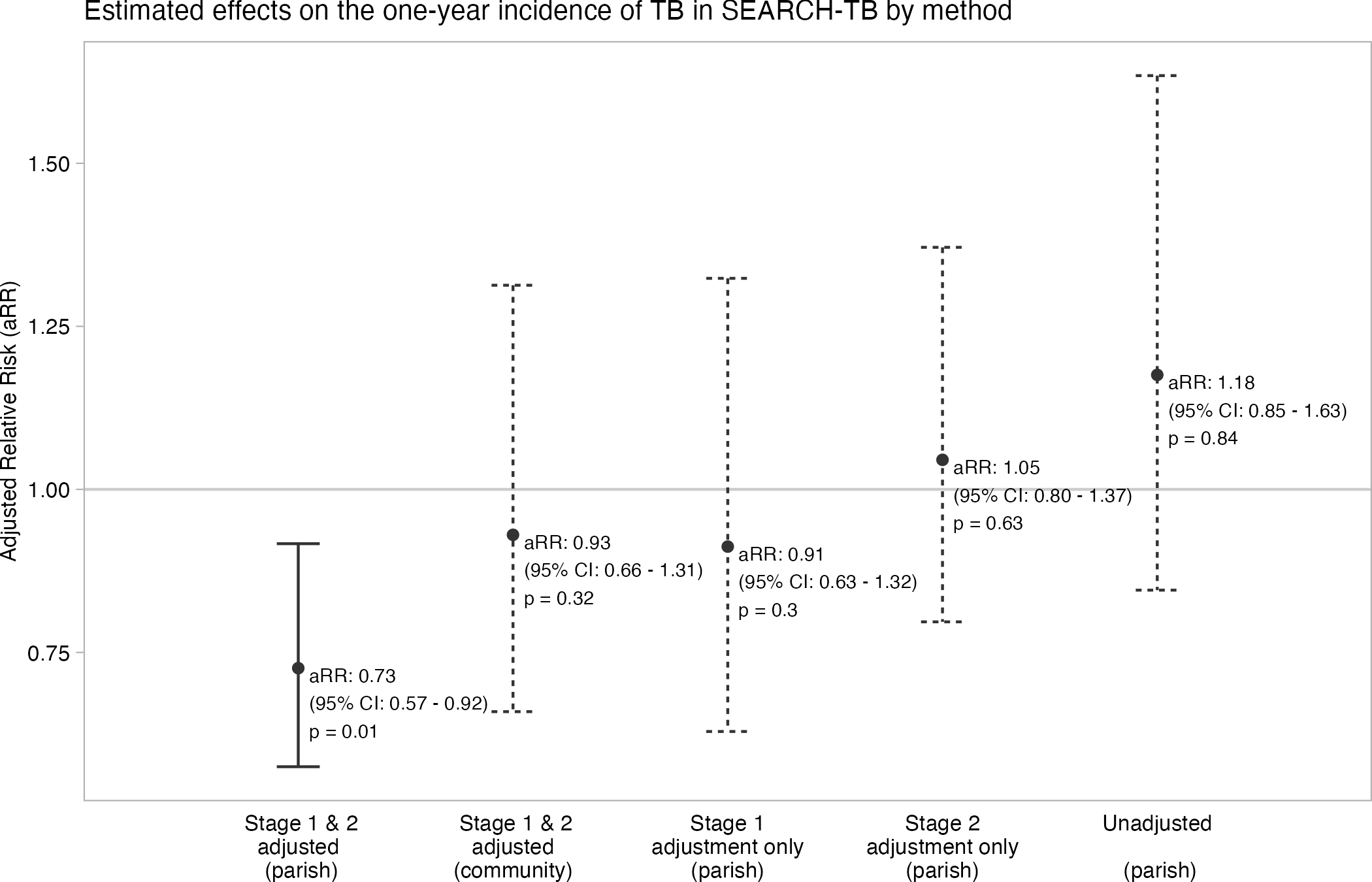}
    \caption{Comparative results under different sets of assumptions using real data from the SEARCH-TB study on incident tuberculosis (TB) infection.  
    The primary analysis, adjusting in both Stage 1 and Stage 2 and considering the parish (a large sub-unit of the community) to be the conditionally independent unit, is shown first. See Table 1 of the Supplementary Materials for additional information.}
    \label{results_fig}
\end{center}
\end{figure}

\clearpage
\newpage

Supplemental materials for ``Blurring cluster randomized trials and observational studies: Two-Stage TMLE for sub-sampling, missingness, and few independent units"

\clearpage
\newpage

\printbibliography

\clearpage
\newpage

\appendix

\section{Derivations of the Stage 1 identifiablity results}
\label{appendix:derivations}

\subsection{Stage 1 denominator $\mathbb{P}(Y_0^*=1)$}
\label{appendix:derivations:denominator}

Recall that $S$ is an indicator of being sampled for participation in the sub-study, $L_0$ are individual-level covariates, $\Delta_0$ is an indicator of individual measurement at baseline, $Y^*_0$ is an underlying indicator of having the outcome at baseline, and $Y_0 = \Delta_0 \times Y^*_0$ an indicator that the individual was measured and had the outcome at baseline. Our goal is to estimate the baseline outcome prevalence $\mathbb{P}(Y^*_0 = 1)$ under a hypothetical intervention where all participants were included $(S = 1)$ and all had their outcome  measured at baseline $(\Delta_0 = 1 \mid S=1)$. Under these following assumptions, together with sufficient data support (i.e., positivity), we can identify the target causal parameter $\mathbb{P}(Y^*_0 = 1)$  as
\begin{equation}
\label{eq:one}
\begin{aligned}
    \mathbb{P}(Y^*_0 = 1) &= \sum_{l_0} \mathbb{P}(Y^*_0=1 \mid  L_0 = l_0) \mathbb{P}(L_0 = l_0) \\
    &\text{by } Y_0^* \perp \!\!\! \perp S \mid L_0 \text{:}\\
     &= \sum_{w} \mathbb{P}(Y^*_0=1 \mid  S = 1, L_0 = l_0) \mathbb{P}(L_0 = l_0) \\
    &\text{by } Y_0^* \perp \!\!\! \perp \Delta_0 \mid S=1, L_0 \text{:}\\
    &= \sum_{l_0} \mathbb{P}(Y^*_0 \mid \Delta_0 = 1, S = 1,L_0 = l_0) \mathbb{P}(L_0 = l_0)\\
    & \text{since }Y_0^* = Y_0 \text{ when } \Delta_0 = 1 \text{:}\\
    &= \sum_{l_0} \mathbb{P}(Y_0=1 \mid \Delta_0 = 1, S = 1, L_0 = l_0) \mathbb{P}(L_0 = l_0)\\
    &=\mathbb{E} \left[ \mathbb{E} \left(Y_0 \mid \Delta_0 = 1, S = 1, L_0 \right) \right].
\end{aligned}
\end{equation}
Throughout the summation generalizes to an integral for continuous-valued variables $L_0$. For the statistical estimand to be well-defined, we also need positivity: 
$\mathbb{P}(S=1 \mid L_0 = l_0) > 0$ and $\mathbb{P}(\Delta_0 = 1 \mid S = 1, L_0 = l_0) > 0$ for all possible values $l_0 \in L_0$.

\subsection{Stage 1 numerator $\mathbb{P}(Y_1^* = 1, Y_0^* = 0)$}
\label{appendix:derivations:numerator}

In addition to the notation in Appendix \ref{appendix:derivations:denominator}, recall that $L_1$ denotes the post-baseline covariates, $\Delta_1$ is an indicator of outcome measurement at follow-up, $Y^*_1$ is the underlying indicator of having the outcome at follow-up, and $Y_1 = \Delta_1 \times Y^*_1$ an indicator of being measured and having the outcome at follow-up. Our goal is to identify the proportion of individuals who have the outcome at follow-up and were at risk at baseline: $\mathbb{P}(Y_1^* = 1, Y_0^* = 0)$. To do so, we utilize and extend the approach of Balzer et al. \cite{balzer_far_2020}. For simplicity of presentation,  we define the underlying indicator of the joint outcome of interest as
$Z^* \equiv \mathbb{I}(Y^*_1 = 1, Y^*_0 = 0)$ and 
 its observed analog as $Z = \mathbb{I}(Y_1 = 1, Y_0 = 0)$.
Under the following causal assumptions, together with sufficient data support (i.e.,  positivity), we can identify $\mathbb{P} (Z^*  = 1)$ as

%

\begin{align*}
\mathbb{P} (Z^* & = 1) \\
=& \sum_{l_0} \mathbb{P}(Z^*=1 \mid  L_0= l_0)\mathbb{P}(L_0=l_0)\\
    \text{by }& Z^* \perp \!\!\! \perp S \mid L_0 \text{ and }  Z^* \perp \!\!\! \perp \Delta_0 \mid S = 1, L_0 \\
=& \sum_{l_0} \mathbb{P}(Z^*=1 \mid \Delta_0=1, S=1, L_0=l_0)\mathbb{P}(L_0=l_0)\\
=& \sum_{l_0}  \sum_{y_0} \sum_{l_1} \mathbb{P}(Z^*=1 \mid L_1=l_1, Y_0=y_0, \Delta_0=1, S=1, L_0=l_0)  \times \\
    & \quad \quad \mathbb{P}(L_1=l_1, Y_0= y_0 \mid \Delta_0=1, S=1, L_0=l_0) \mathbb{P}(L_0=l_0)\\
\text{by } & Z^* = 0 \text{ when } Y_0 = 1 
\\
=& \sum_{l_0}  \sum_{l_1} \mathbb{P}(Z^*=1 \mid L_1=l_1, Y_0=0, \Delta_0=1, S=1, L_0=l_0)  \times  \\
    & \quad \quad \mathbb{P}(L_1=l_1, Y_0= 0 \mid \Delta_0=1, S=1, L_0=l_0)  \mathbb{P}(L_0=l_0)\\
 \text{by } & Z^* \perp \!\!\! \perp \Delta_1 \mid L_1,  Y_0 = 0, \Delta_0 = 1,S=1, L_0  \\
=& \sum_{l_0}  \sum_{l_1} \mathbb{P}(Z^*=1 \mid \Delta_1=1, L_1=l_1, Y_0=0, \Delta_0=1, S=1, L_0=l_0)  \times  \\
    & \quad \quad \mathbb{P}(L_1=l_1, Y_0= 0 \mid \Delta_0=1, S=1, L_0=l_0)  \mathbb{P}(L_0=l_0)\\
=& \sum_{l_0}  \sum_{l_1} \mathbb{P}(Y_1 =1 \mid \Delta_1=1, L_1=l_1, Y_0=0, \Delta_0=1, S=1, L_0=l_0)  \times  \\
    & \quad \quad \mathbb{P}(L_1=l_1 \mid Y_0= 0, \Delta_0=1, S=1, L_0=l_0) \mathbb{P}(Y_0= 0 \mid \Delta_0=1, S=1, L_0=l_0) \mathbb{P}(L_0=l_0)\\
&= \mathbb{E} \left[ \mathbb{E} \{ \mathbb{E} (Y_1 \mid \Delta_1 = 1, L_1, Y_0=0,\Delta_0 = 1, S = 1, L_0) \mid \Delta_0 = 1, S = 1, L_0\} \right]
\end{align*}
As before, the summation generalizes to an integral for continuous-valued covariates and we need the corresponding positivity assumption to hold.


\section{Step-by-step implementation of TMLE}
\label{app:tmle}

\subsection{Stage 1 denominator $\psi_{den}^c$}

Within each cluster, we need to estimate the baseline prevalence of the observed outcome, adjusted for differences between participants
with measured versus missing outcomes:
$$
\psi_{den}^c = \mathbb{E} \left[ \mathbb{E} \left(Y_0 \mid \Delta_0 = 1, S = 1, L_0 \right) \right]
$$
where superscript $c$ emphasizes this is a cluster-level endpoint, shared by all members of that cluster.
As before, $Y_0$ is an indicator of being measured and having the outcome at baseline, $\Delta_0$ is an indicator of measurement, $S$ an indicator of being sampled, and $L_0$ is the set of baseline covariates for the individual and may include characteristics of household members or other contacts.

For each cluster $i=\{1,\ldots, N\}$, the process for estimating $\psi_{den,i}^c$ using TMLE is as follows:

\begin{enumerate}
    \item Subset the data to the $r=\{1,\ldots, R_i\}$ individuals in that cluster.
    \item Among those sampled and measured ($\Delta_0 = 1, S = 1$), flexibly estimate the expectation of the outcome $Y_0$ given the adjustment covariates $L_0$, using Super Learner, an ensemble method to combine predictions from a diverse set of machine learning algorithms \cite{laan_super_2007}.
    \item Use the estimated function from Step 2 to predict the outcome for all individuals, measured or not, taking into account their covariates: $\hat{\mathbb{E}}(Y_0 \mid \Delta_0 = 1, S = 1, L_{0,r})$ for $r = 1, \dots, R_i$.
    \item Flexibly estimate the measurement mechanism, which is the conditional probability of sampling and measurement given the adjustment covariates $\mathbb{P}(\Delta_0 = 1, S = 1 | L_0)$, with Super Learner.
    \item Use the estimated function from Step 4 to predict the ``propensity'' for measurement $\hat{\mathbb{P}}(\Delta_0 = 1, S = 1 | L_{0,r})$ for all individuals $r = 1, \dots, R_i$. Then calculate the ``clever covariate'' $\hat{H_r} = \frac{\mathbb{I}(\Delta_{0,r} = 1, S_r = 1)}{\hat{\mathbb{P}}(\Delta_0 = 1, S = 1 | L_{0,r})}$  for all individuals $r = 1, \dots, R_i$.
    \item Target the initial outcome predictions $\hat{\mathbb{E}}(Y_0 \mid \Delta_0 = 1, S = 1, L_{0})$:
    \begin{enumerate}
        \item Run an intercept-only logistic regression on the observed outcomes $Y_0$, including the $logit$ of the initial outcome predictions as an offset and the clever covariate $\hat{H}$ as the weight. 
       \item Use the prior regression to obtain targeted predictions of the outcome for all individuals:  $\hat{\mathbb{E}}^*(Y_0 \mid \Delta_0 = 1, S = 1, L_{0,r})$ for $r=\{1,\ldots, R_i\}$. This corresponds to adding the estimated intercept, denoted $\hat{\epsilon}$, to the the $logit$ of the initial  predictions and  transforming back to the original scale.
        \begin{align*}
        \hat{\mathbb{E}}^*(Y_0 \mid \Delta_0 = 1,& S = 1, L_{0}) =logit^{-1} \left[ \hat{\epsilon} + logit\{\hat{\mathbb{E}}(Y_0 \mid \Delta_0 = 1, S = 1, L_{0}) \} \right].
        \end{align*}
    \end{enumerate}
    \item Average the targeted predictions to estimate the baseline outcome prevalence, accounting for sub-sampling and missingness:
    $$
    \hat{\psi}_{den,i}^c = \frac{1}{R_i} \sum_{r=1}^{R_i} \hat{\mathbb{E}}^*(Y_0 \mid \Delta_0 = 1, S = 1, L_{0,r})
    $$
\end{enumerate}
We note that in Step 4, estimation of the measurement mechanism $\mathbb{P}(\Delta_0 = 1, S = 1 | L_0)$ can follow the factorization $\mathbb{P}(\Delta_0 = 1\mid S = 1, L_0) \times \mathbb{P}(S = 1 \mid L_0)$, but this is not necessary. If the partition is, instead, considered to be the effective independent unit, TMLE can be implemented analogously to estimate partition-level parameter $\psi_{den,k}^p$ in each partition $k=\{1,\ldots,K\}$.

\subsection{Stage 1 numerator $\psi_{num}^c$}

Recall the causal parameter for the numerator $\mathbb{P}(Y_1^* = 1, Y_0^* = 0)$ corresponds to hypothetical intervention  to 
(1) include all cluster members in the sample and ensure their baseline outcome measurement, 
and (2) ensure follow-up outcome measurement among all known to be at-risk at baseline. 
Interventions that are responsive to the participant's past are sometimes called ``dynamic treatment regimes''
\cite{hernan_comparison_2006, laan_causal_2007, robins_estimation_2008}. We can formally define our dynamic interventions as the following functions: \begin{itemize}
    \item $d_0(\cdot)$: set $S=1$ and $\Delta_0=1$ for all
    \item $d_1(Y_0, \Delta_0)$: if $Y_0=0$ and $\Delta_0=1$, set $\Delta_1=1$; otherwise, set $\Delta_1=0$
\end{itemize}
For ease of presentation, we simplify the notation for these functions to $d_0$ and $d_1$, respectively.
Under the previously stated assumptions in Appendix \ref{appendix:derivations:numerator} \cite{robins_new_1986}, the causal parameter corresponding to these dynamic interventions is identified as
$\psi_{num}^c=\mathbb{E} \left[ \mathbb{E} \{ \mathbb{E} (Y_1 \mid \Delta_1 = 1, L_1, Y_0=0,\Delta_0 = 1, S = 1, L_0) \mid \Delta_0 = 1, S = 1, L_0\} \right]$.

For ease of presentation, let $X_0$ denote the baseline covariates; $A_0=\{S, \Delta_0\}$ denote the baseline intervention variables; $X_1=\{Y_0,L_1\}$ denote the intermediate non-intervention variables, and $A_1=\Delta_1$ denote the follow-up intervention variable. As before, the observed outcome at follow-up is denoted $Y_1$.
 Then our identifiability result can be re-written as 
$$\psi_{num}^c=\mathbb{E} \left[ \mathbb{E} \{ \mathbb{E} (Y_1 \mid A_1=d_1, X_1, A_0=d_0, X_0) \mid A_0=d_0, X_0\} \right]$$
Now we outline the implementation of longitudinal TMLE for $\psi_{num}^c$ using the iterated conditional expectation formulation \cite{Bang&Robins05, vanderLaan2012towertmle}.

\begin{enumerate}
\item  Subset the data to the $r=\{1,\ldots, R_i\}$ individuals in  cluster $i$.
\item Flexibly estimate the inner-most conditional outcome expectation $\mathbb{E} (Y_1 \mid A_1, X_1, A_0, X_0)$ with Super Learner. 
Then predict the outcome for all individuals, under the dynamic regime of interest: $\hat{\mathbb{E}}(Y_1 \mid A_1=d_1, X_{1,r}, A_0=d_0, X_{0,r})$ for $r=\{1, \ldots, R_i\}$.
\item Target these initial predictions using information in the measurement mechanism.
    \begin{enumerate}
    \item Flexibly estimate the measurement mechanism with Super Learner and calculate the ``clever covariate'': 
$$ \hat{H}_{1,r} = \frac{\mathbb{I}(A_{0,r}=d_0, A_{1,r}=d_1)}{\hat{\mathbb{P}}(A_0 \mid X_{0,r})\times \hat{\mathbb{P}}(A_1 \mid X_{1,r}, A_{0,r}, X_{0,r})}$$ for all individuals $r = 1, \dots, R_i$.
\item Run an intercept-only logistic regression on the observed outcome $Y_1$  using the $logit$ of the initial outcome predictions $\hat{\mathbb{E}}(Y_1 \mid A_1=d_1, X_{1}, A_0=d_0, X_{0})$ as  offset and the clever covariate $\hat{H}_{1}$ as the weight.
    \item  Use this regression to obtain targeted predictions of the outcome for all individuals under the regime of interest:  $\hat{\mathbb{E}}^*(Y_1 \mid A_1=d_1, X_{1,r}, A_0=d_0, X_{0,r})$ for $r=\{1, \ldots, R_i\}$.
\end{enumerate}
    
\item Consider these targeted predictions $\hat{\mathbb{E}}^*(Y_1 \mid A_1=d_1, X_{1}, A_0=d_0, X_{0})$ to be psuedo-outcomes and use Super Learner to flexibly estimate their relation to the intervention and covariates at baseline $(A_0, X_0)$. Then predict a new psuedo-outcome for all individuals, under the treatment regime of interest: 
$\hat{\mathbb{E}} \{ \hat{\mathbb{E}}^*(Y_1 \mid A_1=d_1, X_{1,r}, A_0=d_0, X_{0,r}) \mid A_0=d_0, X_{0,r}  \} $ for $r=\{1, \ldots, R_i\}$. 

\item Target these initial predictions using information in the measurement mechanism.
    \begin{enumerate}
    \item Calculate a new  ``clever covariate'': 
$ \hat{H}_{0,r} = \frac{\mathbb{I}(A_{0,r}=d_0)}{\hat{\mathbb{P}}(A_0=d_0 \mid X_{0,r})}$ for all individuals $r = 1, \dots, R_i$.
\item Run an intercept-only logistic regression of the ``outcome''
$\hat{\mathbb{E}}^*(Y_1 \mid A_1=d_1, X_{1}, A_0=d_0, X_{0})$
using the $logit$ of the initial estimates $\hat{\mathbb{E}} \{ \hat{\mathbb{E}}^*(Y_1 \mid A_1=d_1, X_{1}, A_0=d_0, X_{0}) \mid A_0=d_0, X_{0}  \}$ as offset and with weights $\hat{H}_{0}$.
    \item  Use this regression to obtain targeted predictions of the ``outcome'' for all individuals under the treatment regime of interest:  $\hat{\mathbb{E}}^* \{ \hat{\mathbb{E}}^*(Y_1 | A_1=d_1, X_{1,r}, A_0=d_0, X_{0,r}) \mid A_0=d_0, X_{0,r}  \} $ for $r=\{1, \ldots, R_i\}$ 
    \end{enumerate}
    
\item Average these final predictions to standardize with respect to the baseline covariate distribution:
    $$\hat{\psi}_{num,i}^c=\frac{1}{R_i} \sum_{r=1}^{R_i}
 \hat{\mathbb{E}}^* \{ \hat{\mathbb{E}}^*(Y_1 \mid A_1=d_1, X_{1,r}, A_0=d_0, X_{0,r}) \mid A_0=d_0, X_{0,r}  \} .$$
\end{enumerate}

Worked examples of estimation dynamic interventions with longitudinal TMLE are provided in Schomaker et al. \cite{Schomaker2019} and Montoya et al. \cite{Montoya2022}.
Software packages, such as \texttt{ltmle}, are readily available and can easily accommodate censoring (i.e., measurement indicators) as well as deterministic knowledge (e.g., $Y_1=1$ if $Y_0=1$) \cite{lendle_ltmle_2017}. Code for the applied example is available at \url{https://github.com/joshua-nugent/search-tb}.
As before, if the partition is, instead, considered to be the effective independent unit, TMLE can be implemented analogously to estimate partition-level parameter $\psi_{num,k}^p$ in each partition $k=\{1,\ldots,K\}$.

\newpage

\begin{table}[!h]
    \centering
    \def\arraystretch{1.5}
    \begin{tabularx}{\textwidth}{Y Y l}
        \textbf{Estimator}  & \textbf{Key assumptions} & \makecell{\textbf{Risk ratio}\\\textbf{estimate (95\% CI)}}\\
        \hline
Stage 1 adjusted; Stage 2 adjusted; parish as  the independent unit (Primary analysis) &
        Individual-level outcomes are missing at random (MAR) given household HIV status, age and mobility.  Parishes-level outcomes are conditionally independent given the prevalence of HIV and prevalence of alcohol use. & 0.73 (0.57 - 0.92)\\
Stage 1 adjusted; Stage 2 adjusted; community as the independent unit$^*$  &
    Individual-level outcomes are MAR given household HIV status, age and mobility.
       & 0.93 (0.66 - 1.31)\\
Stage 1 adjusted; Stage 2 unadjusted; parish as the independent unit  &
    Individual-level outcomes are MAR given household HIV status, age and mobility.  Parishes-level outcomes are (marginally) independent. 
        & 0.91 (0.63 - 1.32)\\
Stage 1 unadjusted; Stage 2 adjusted; parish as the independent unit &
        Individual-level outcomes are missing completely at random (MCAR). Parishes-level outcomes are conditionally independent given the prevalence of HIV and prevalence of alcohol use.
    & 1.05 (0.80 - 1.37)\\
Stage 1 unadjusted; Stage 2 unadjusted; parish as the independent unit  & 
    Individual-level outcomes are MCAR. Parishes-level outcomes are (marginally) independent. 
&  1.18 (0.85 - 1.63)\\
        \hline
\multicolumn{3}{l}{\begin{footnotesize} $^*$Stage 2  adjustment covariates selected through Adaptive Prespecification to maximize empirical efficiency \cite{balzer_adaptive_2016}. \end{footnotesize}}
    \end{tabularx}
    \caption{Comparison of the results under different sets of assumptions using real data from the SEARCH-TB on incident tuberculosis (TB) infection in rural Uganda.}
    \label{tab:sens}
\end{table}

\end{document}